\documentclass[aps,prl,twocolumn,floatfix]{revtex4}
\usepackage{amssymb}

\usepackage{epsfig}
\usepackage{amsmath}

\newcommand{\beq}[1]{\begin{eqnarray} \label{#1}}
\newcommand{\eeq}{\end{eqnarray}}

\begin{document}

\title{Absorption in the fractional quantum Hall regime: trion dichroism and
spin polarization}
\author{J. G. \surname{Groshaus} $^{1,2,3}$}
\email{jgg@phys.columbia.edu}
\author{P. \surname{Plochocka-Polack} $^{1}$}
\author{M. \surname{Rappaport} $^1$}
\author{V. \surname{Umansky} $^{1}$}
\author{I. \surname{Bar-Joseph} $^{1}$}
\affiliation{$^1$ Department of Condensed Matter Physics, The Weizmann Institute of
Science, Rehovot, Israel.}
\author{B. S. \surname{Dennis} $^2$}
\author{L. N \surname{Pfeiffer} $^2$}
\author{K. W. \surname{West} $^{2}$}
\author{Y. \surname{Gallais} $^{3}$}
\author{A. \surname{Pinczuk} $^{2,3}$}
\affiliation{$^2$ Bell Laboratories, Lucent Tech. Murray Hill, NJ 07974.}
\affiliation{$^3$ Department of of Physics and of Appl. Physics and Appl. Mathematics,
Columbia University, New York, NY 10027}
\date{\today }

\begin{abstract}
We present measurements of optical interband absorption in the fractional
quantum Hall regime in a GaAs quantum well in the range $0<\nu \leq 1$. We
investigate the mechanism of singlet trion absorption, and show that its
circular dichroism can be used as a probe of the spin polarization of the
ground state of the two-dimensional electron system (2DES). We find that at $%
\nu \leq 1/3$ the 2DES is fully spin-polarized. Increasing the filling
factor results in a gradual depolarization, with a sharp minimum in the
dichroism near $\nu =2/3$. We find that in the range $0.5\leq \nu <0.85$ the
2DES remains partially polarized for the broad range of magnetic fields from
2.75 to 11 Tesla. This is consistent with the presence of a mixture of
polarized and depolarized regions.
\end{abstract}

\maketitle


The electron-electron Coulomb interaction plays an important role in
determining the spin polarization ($\mathcal{P}$) of a two-dimensional
electron system (2DES) in a perpendicular magnetic field $B$. The effect of
the interactions on the spin is most readily seen in in the filling factor
range $\nu \leq 1$, where the level degeneracy exceeds the number of
electrons. The exchange part of the Coulomb interactions favors a
ferromagnetic state. However, a many-body wavefunction in which all
electrons have the same spin is restricted by the Pauli principle and may
not constitute in general an optimal spatial distribution that minimizes the
total Coulomb energy. Inclusion of components of the opposite spin in the
wavefunction opens a larger phase space for the electrons and may result in
a lower Coulomb repulsion. Thus, the ground state of the 2DES in this regime
is determined by the competition between the gain in Coulomb energy and the
cost in Zeeman energy.

Several methods have been used to determine the 2DES spin polarization in
the fractional quantum Hall regime. The direct method is measuring the shift
in the nuclear magnetic resonance (NMR) caused by the 2DES magnetization.
Indeed, NMR measurements have yielded quantitative measurements of $\mathcal{%
P}$ throughout a broad range of filling factors \cite{Khandelwal98, Freytag}%
. Transport experiments have also been used and were successful in
identifying transitions in the spin polarization, although are less
effective in providing a quantitative value of $\mathcal{P}$ \cite%
{Perspectives, Smet}.

An alternative approach for measuring $\mathcal{P}$ is using optical
spectroscopy: photoluminescence \cite{Kukushkin}, reflectivity \cite%
{Chughtai2002}, and absorption spectroscopy \cite{Aifer96}. In these
techniques, the occupation of each spin levels are obtained from
measurements of the circular dichroism of the interband transitions. In
quantum wells (QW), however, establishing the relation between the optical
oscillator strengths (OS) and $\mathcal{P}$ presents a major difficulty. As
a result of the strong interaction between the photo-created valence hole
and the electrons, the spectrum in QWs is dominated by resonance peaks,
associated with the neutral exciton, and charged excitons (trions). A
further complication is introduced by the fact that in a trion, the two
electrons can form a singlet or triplet wavefunctions \cite{Yusa2001,Wojs}.
Hence, it is essential to take into account the nature and symmetry of these
excitonic states \cite{Groshaus2004}.

In this paper we show that the 2DES spin polarization can be quantitatively
determined by measuring the circular dichroism of the singlet trion
absorption, $\mathcal{D}_{T}$, defined as $(I_{T_{\downarrow
}}-I_{T_{\uparrow }})/(I_{T_{\downarrow }}+I_{T_{\uparrow }})$, where $%
I_{T_{\uparrow ,\downarrow }}$ are the OS of the singlet trion peaks at the $%
\sigma^+$ $(\sigma^-)$ circular light polarization. We apply this approach
to back-gated quantum well samples, in which the 2DES density can be
continuously varied. By determining the electron density dependence of $%
\mathcal{D}_{T}$\ at different magnetic fields we map the spin polarization
of a 2DES in the range $\nu <1$. We find that the 2DES is spin polarized at $%
\nu \leq 1/3$, exhibits a gradual depolarization at $\nu >1/3 $, leveling at
$\mathcal{P}\approx 1/2$, exhibit a sharp minimum near $\nu=2/3$ and
re-polarizes towards $\nu=1$. We find that the results are largely
independent on the magnetic field in the range studied. This is consistent
with the proposed presence of magnetic domains \cite{Murthy2000,
Smet,OStern2004, Schulze2004b}.

The experiments were done in a dilution fridge with optical windows at a
base temperature of 70 mK, and a magnetic field applied along the growth
axis of the wafer. Two experimental techniques were implemented to measure
the absorption spectrum. The first is based on measuring the photocurrent
(PC) flowing between the 2DES and a back-gate. The sample is illuminated by
a tunable dye laser with power densities of $\sim 5\times 10^{-4}$ W/cm$^{2}$
through a circular polarizer, and the PC is measured as a function of the
photon energy \cite{Groshaus2004}. In the second technique we incorporate a
Bragg mirror between the 2DES and the back gate. White light at normal
incidence ($<1$ nW) is back reflected from this mirror, and passes twice
through the QW. Thus, the reflected spectra, divided by the signal at
energies below the gap, yields the transmission spectrum.

Two samples were measured: Sample 1 was used for the PC measurements and
consists of a single 20 nm GaAs/Al$_{0.12}$Ga$_{0.82}$As modulation-doped QW
grown on top of a 1.7 $\mu $m Al$_{0.3}$Ga$_{0.7}$As barrier layer,
separating it from the back-gate layer. Sample 2 was used for the
transmission measurements and includes in addition a Bragg reflector,
consisting of 20 pairs of AlAs and Al$_{0.3}$Ga$_{0.60}$As layers. The
wafers were processed to a mesa structure with selective ohmic contacts to
the 2DES and to the back-gate. Applying a voltage between the 2DES and the
back-gate we could tune the electron density continuously in the range $%
1\times 10^{10}-1\times 10^{11}$ cm$^{-2}$ in sample 1 and $5\times
10^{10}-3\times 10^{11}$ in sample 2.

\begin{figure}[tbp]
\epsfig{figure=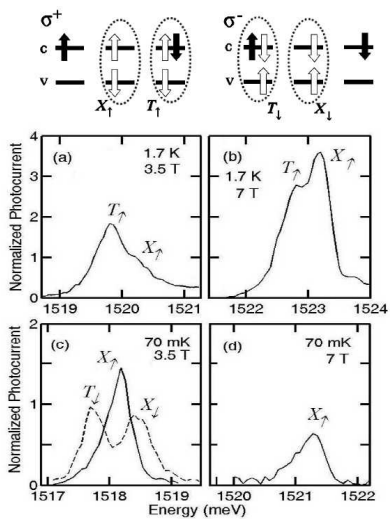, width=0.6\linewidth, clip=}
\caption{(upper scheme) The mechanisms for absorption in quantum wells for
both light polarizations. Horizontal lines represent orbitals in the
conduction and valence band, \emph{c} and \emph{v}. Black arrows are
preexisting electrons in the 2DES, and white arrows are photo-created
electron-hole pairs. A photo-excited pair overlapping an existing electron,
binds it forming a singlet trion; (a,b,c,d) Photocurrent (PC) spectra at $%
\protect\nu =1/3$ for absorption into the lower and upper Zeeman level
(solid and dashed lines lines, respectively. For clarity, the spectra at
each gate voltage is normalized by the PC efficiency, namely, we divide by
the the PC signal at $B=0$ at photon energies well above the Fermi energy,
where the absorption is assumed to be independent of the voltage.}
\label{fig:1}
\end{figure}

The solid lines in Fig. 1 depict four photocurrent spectra at the $\sigma
^{+}$ light polarization (absorption into the lower Zeeman level) for two
magnetic fields, $3.5$ and $7$ T, and two temperatures, $1.7$ K and $70$ mK,
at approximately constant $\nu\approx1/3$. It is seen that the spectra at $%
1.7$ K consist of two peaks: $T_{\uparrow }$ and $X_{\uparrow }$. However,
as the sample is cooled to $70$ mK - the $T_{\uparrow }$ peak completely
disappears.

This striking behavior can be understood by noting the \textit{singlet}
nature of the $T_{\uparrow ,\downarrow }$ trionic states. The symmetrical
nature of the spatial wavefunction of the electrons in the trion implies
that it is formed when an electron is photo-created in an orbital where an
electron with \textit{opposite} spin is already present. For example, the $%
T_{\uparrow }$ transition occurs when a $\sigma ^{+}$ photon creates an $%
\uparrow $-electron (in the lower Zeeman level) in spatial orbitals which
are occupied by $\downarrow $-electrons (see the upper scheme of Fig. 1).
Thus, the probability of creating the $T_{\uparrow }$ state will depend not
only on the phase space available for absorption into $\uparrow$-states, but
also on the total number of $\downarrow $-electrons. The $X$ peaks, on the
other hand, are formed when an electron is photo-created in unoccupied
orbitals \cite{comment-Triplet}.

Within this picture, the spectrum is readily explained: at $1.7$ K the 2DES
is not spin-polarized, and hence $\downarrow $- electrons are present at the
upper Zeeman level. These $\downarrow $- electrons can pair with the
photo-created $\uparrow $-electron giving rise to the $T_{\uparrow }$ peak.
As the sample is cooled to 70 mK at $\nu =1/3$, the 2DES becomes
spin-polarized, and the absence of $\downarrow $-electrons results in a the
quenching of the $T_{\uparrow }$ peak. This interpretation is confirmed by
the fact that in absorption to the upper Zeeman level (Fig. 1c, dashed
line), the $T_{\downarrow }$ peak persists at low temperatures: there are
always $\uparrow $-electrons to pair with the photo-created $\downarrow $%
-electron. The resulting dichroism, $\mathcal{D}_{T}$, is equal to 1.

Increasing the density above $\nu =1/3$ we find that the $T_{\uparrow }$
peak, which was absent at $\nu \leq 1/3$, gradually appears and gains
strength, indicating a loss of spin polarization. This is demonstrated in
Fig. 2a, which shows the $\sigma ^{+}$ spectra at $T=70$ mK and constant
magnetic field, $B=4$ T, for $0.3\lesssim \nu \lesssim 0.7$. It is seen that
around $\nu \approx 2/3$ the $T_{\uparrow }$ peak is already well developed.
Figure 2b shows spectra taken at $\nu \approx 2/3$ at several magnetic
fields, $2.75<B<5$ T. It is seen that the $T_{\uparrow }$ peak exists in all
spectra, and becomes the dominant one as the field increases. These
indications for depolarization above $\nu =1/3$ are observed over a
relatively broad magnetic field range. Unfortunately, however, at high
fields ($B\geq7$ T) the spectral lines in this sample are broadened near $%
\nu =1/3$ and the trion peak height cannot be reliably resolved.

\begin{figure}[tbp]
\epsfig{figure=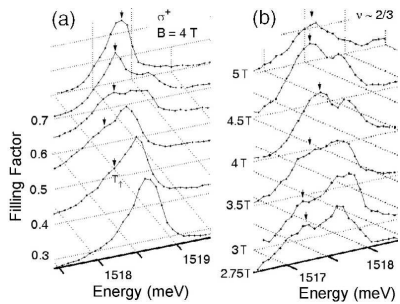, width=0.6\linewidth, clip=}
\caption{Normalized photocurrent spectra at the $\protect\sigma ^{+}$ light
polarization: (a) at constant field ($B=4$ T) for several $\protect\nu $. As
$\protect\nu $ increases above $1/3$, the $T_{\uparrow }$-peak, marked by
the arrows, appears and increases. (b) At constant $\protect\nu $ ($\sim 2/3$%
) for several $B$. The $T_{\uparrow }$-peak is present throughout the range,
and increases with the electron density.}
\label{fig:Nuscan}
\end{figure}

To obtain the single trion circular dichroism we determine $I_{T_{\uparrow
}} $ and $I_{T_{\downarrow }}$ by taking the area under each of the
corresponding $T$ peaks. Figure 3 shows $\mathcal{D}_{T}$ as a function of $%
\nu $ at two magnetic fields, $B=2.75$ and $5$ T. Remarkably, we find a
similar dependence of $\mathcal{D}_{T}$ on $\nu $ throughout this magnetic
field range: $\mathcal{D}_{T}\approx 1$ at $\nu \leq 1/3$, exhibits a fast
decrease at $\nu >1/3$, leveling at $\mathcal{D}_{T}\approx 1/2$, and
increases again to $\mathcal{D}_{T}\approx 1$ at $\nu =1$.

\begin{figure}[tbp]
\epsfig{figure=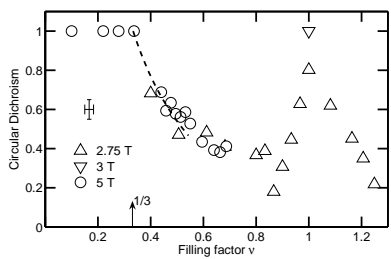, width=0.6\linewidth, clip=}
\caption{The circular dichroism of the singlet trion absorbtion, $\mathcal{D}%
_{T}$. The dashed line is the theoretical depolarization caused by a
quasiparticle of $\mathcal{S}=0.23$. The error bar shows typical errors in $%
\protect\nu $ and $\mathcal{D}_{T}$. The values shown for $\protect\nu >1$
are given by $(2-\protect\nu )/\protect\nu \ \mathcal{D}_{T}$) \protect\cite%
{Comment nu>1}. }
\label{fig:polall}
\end{figure}

To interpret these results, we need to relate the trion absorption, $%
I_{T_{\uparrow ,\downarrow }}$, to the spin polarization. Since the
photo-created electron needs to pair with an opposite spin, we model the OS
of each $T$ peak as proportional to the number of unpaired electrons with
opposite spin, i.e.

\begin{eqnarray}
I_{T_{\uparrow }} &=&Cf_{\downarrow }N_{\downarrow }=Cf_{\downarrow }N_{e}%
\frac{1}{2}(1-\mathcal{P})  \label{Tmodel} \\
I_{T_{\downarrow }} &=&Cf_{\uparrow }N_{\uparrow }=Cf_{\uparrow }N_{e}\frac{1%
}{2}(1+\mathcal{P}).  \notag
\end{eqnarray}%
Here $N_{\downarrow (\uparrow )}$ are the number of $\downarrow (\uparrow )$%
-electrons, $\mathcal{P}\equiv (N_{\downarrow }-N_{\uparrow })/N_{e}$, and $%
C $ is a proportionality constant, which may depend on $B$ and $\nu $, and
includes the PC efficiency. The factor $f_{\downarrow (\uparrow )}$ is the
fraction of $\downarrow (\uparrow )$ - electrons that are \emph{unpaired},
i.e. that occupy orbitals where the state of opposite spin is empty. In an
uncorrelated gas in the lowest Landau level approximation, this factors are
proportional to the single-particle phase space available, i.e. $%
f_{\downarrow }=1-N_{\uparrow }/N_{\phi }$. In a strongly correlated 2DES,
the Coulomb repulsion tends to minimize the number of paired electrons.
Therefore, for $\nu \leq 1$, we shall assume that in the ground state all
the electrons are unpaired, i.e. $f_{\downarrow ,(\uparrow )}=1$. It can be
seen from Eq. \ref{Tmodel} that both $I_{T_{\uparrow ,\downarrow }}$ are
expected to increase with $N_{e}$. Figure 2b demonstrates this for the $%
T_{\uparrow }$ peak: the increase of the peak with magnetic field at
constant $\nu $ is due to the increased electron density. The $T_{\downarrow
}$ peak increases accordingly (not shown).

The relation between the spin polarization, $\mathcal{\ P}$, and dichroism $%
\mathcal{D}_{T}$ turns out to be surprisingly simple. Using Eq. \ref{Tmodel}%
, $\mathcal{\ P}$ can be expressed (for $\nu \leq 1$ \cite{Comment nu>1}) as

\begin{equation}
\mathcal{P}=\frac{I_{T_{\uparrow }}-I_{T_{\downarrow }}}{I_{T_{\uparrow
}}+I_{T_{\downarrow }}},  \label{truePol}
\end{equation}%
namely, $\mathcal{P}=\mathcal{D}_{T}$. This model allows us to interpret the
measurements. We focus first on the behavior around $\nu =1/3$ in fig. \ref%
{fig:polall}. As expected \cite{Laughlin83} we find that the 2DES is spin
polarized at $\nu =1/3$ . Near this filling factor the polarization curve is
highly asymmetric: the 2DES is spin polarized at $\nu \lesssim 1/3$, and
gradually depolarizes at $\nu>1/3$.

The depolarization can be expressed in terms of the number of spin flips, $%
\mathcal{S}$, caused by changing the magnetic flux by one quanta: $\mathcal{P%
}(\nu >\nu _{0})=1-2|1/\nu _{0}-1/\nu |\mathcal{S}$, where $\nu _{0}=1/3$. A
similar expression can be written for $\nu <\nu _{0}$, where the change in
spin per flux is denoted as $\mathcal{A}$. The dashed line in Fig. \ref%
{fig:polall} is a fit to the above expression with $\mathcal{S}=0.23$.
Within our experimental error we cannot rule out some depolarization at $\nu
<1/3$, yet an asymmetric behavior is evident, with $\mathcal{A}<<\mathcal{S}$%
. An similar asymmetry in $\mathcal{P}$ around $\nu =1/3$ was also observed
by NMR with $\mathcal{S}\sim 0.1$ around 12 T \cite{Khandelwal98}. The fact
that $\mathcal{S}<1$ is surprising: charge excitation of a quasiparticle
with a reversed spin $1/2$ is expected to yield a net change of spin equal
to $1$, and a more complex quasiparticle, such as a Composite Fermion
Skyrmion \cite{CFSkyrmions} should cause several spin-flips per one flux
quanta, i.e - $\mathcal{S}\geq 1.$ This behavior could be due to a spatial
inhomogeneity in the spin polarization \cite{OStern2004}. Accordingly, the
2DES consists of regions with $\mathcal{P}=0$ and $\mathcal{P}=1$, and the
rate of change in $\mathcal{P}$ represents the gradual change in this
mixture.

Let us turn now to the region around $\nu =2/3$. Figures 4a and 4b show the
measured absorption spectra around this filling factor at $11$ T, at 70 mK
in sample 2, in the lower ($\sigma ^{-}$) and upper Zeeman level ($\sigma
^{-}$). It is clearly seen that the area of the $T_{\downarrow }$ peak ($%
\sigma ^{-}$) is minimal at $\nu =2/3$ (blue line). Simultaneously, the area
of the $T_{\uparrow }$ peak ($\sigma ^{+}$) is maximal at the same filling
factor (even though the peak height at $\nu =0.7$ and $0.75$ are higher),
resulting in a dip in the dichroism. In Fig. 4c we show the dichroism for a
broad range of magnetic fields, $5-11$ T. It is seen that a minimum of the
dichroism always appears in the vicinity of $\nu =2/3$. Within the
uncertainty in filling factor, we are not able to determine in our work if $%
\nu =2/3$ falls inside the dip. It is possible that $\mathcal{P}$ remains
constant for $1/2\geq \nu \leq 2/3$, and then above $2/3$ there is a rapid
depolarization, as was found in Ref. \cite{Freytag}. The dashed line in Fig.
4c shows the depolarization rate with $\mathcal{S}=1$, and it is seen that
it fits well to our data. Such depolarization rate was also found in Ref. %
\cite{Freytag} above $2/3$.

\begin{figure}[tbp]
\epsfig{figure=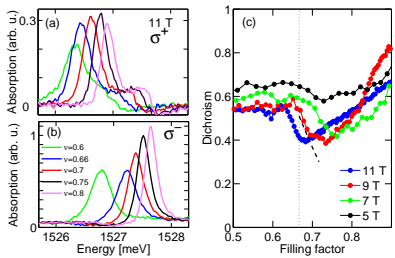, width=0.6\linewidth, clip=}
\caption{(a), (b) The absorption spectrum at at 70 mK and $B = 11$ T for $%
\protect\sigma ^{-}$and $\protect\sigma ^{+}$ polarizaion (absorption to the
upper and lowe Zeeman, respectively) for several filling factors around $%
\protect\nu =2/3$. (c) The circular dichroism of the trion absorption at
several magnetic fields. The dashed line is $\mathcal{P}= \frac{1}{2} -2|1/%
\protect\nu _{0}-1/\protect\nu |\mathcal{S}$, where $\protect\nu _{0}=2/3$,
and $\mathcal{S}=1$.}
\label{fig:two_third}
\end{figure}

We observe that over the range of intermediate filling factors, $0.5\leq \nu
\leq 0.85$, the 2DES remains about half polarized for the whole range of
magnetic fields. In a model of non-interacting composite Fermions (CF) \cite%
{Jain89} $\mathcal{P}$ is expected to be determined by the ratio between
Zeeman, $E_{z}$, and Coulomb energies, $E_{C}$, often represented by the
parameter $\eta =E_{Z}/E_{C}=0.0057\sqrt{B}$ (for perpendicular fields).
Here $E_{Z}=|g|\mu _{B}B$, and $E_{C}=\frac{e^{2}}{\epsilon }\sqrt{eB/(\hbar
c)}$. In this model, at $\nu =2/3$ as $\eta $ is increased across a critical
value $\eta _{c}$, at which the Zeeman energy equals the CF-cyclotron gap,
the polarization is expected to undergo a sharp transition from 0 to 1. At $%
\nu =1/2$, the polarization is expected to raise as $\mathcal{P}=\eta /\eta
_{c}$, until the Zeeman energy equals the CF-Fermi energy, and $\mathcal{P}$
for $\eta >\eta _{c}$. For $\nu =2/3$, the value of $\eta _{c}$ in
literature varies in the range $0.008-0.0185$ \cite{Kukushkin, Perspectives,
Schulze2004a, Freytag, Smet}. For $\nu =1/2$, it was found that $\eta
_{c}=0.022$ \cite{Freytag}, and $0.0166-0.018$ \cite{Kukushkin,
comment-variance}. In our samples we find that the $\eta _{c}$ for all
filling factors between 1/2 and 1 should be higher than our range ($\eta
_{c}>0.0189$).

The fact that at $\nu=2/3$ the spin polarization does not vanish for a large
range of $\eta$ below $\eta_c$ is remarkable, and it has been observed
before for smaller range of $\eta$ \cite{Freytag}. The effect of residual
CF-CF interactions, disorder, spin-orbit coupling predict a partially
polarized phase for a much smaller range of $\eta$ \cite{Eros}. Recently, a
model based on the effect of the lattice potential predicted the existence
of inhomogeneous magnetic domains, resulting in a state with average $%
\mathcal{P}=1/2$ for broader range of $\eta$ \cite{Murthy2000}. Experimental
evidence for the existence of domains was provided by recent spectroscopic
NMR measurements \cite{OStern2004}, by the observation of the breaking in
the selection rules for phonon excitation \cite{Schulze2004b}, and by the
hysteresis of $\eta_c$ in transport \cite{Smet}. Both the rate of
depolarization around $1/3$ and the partial polarization state found around $%
2/3$ and $1/2$ are consistent with this explanation.

This work was supported by the Binational Science Foundation. J.G wishes to
acknowledge fruitful discussions with E. Mariani and Y. Levinson.


\begin{thebibliography}{99}
\bibitem{Khandelwal98} P. Khandelwal \textit{et al.}, \prl{\textbf{81}}, 673
(1998)

\bibitem{Freytag} N. Freytag \emph{et al.}, \prl{\textbf{87}}, 136801
(2001); N. Freytag \emph{et al.}, \prl{\textbf{89}}, 246804 (2002).



\bibitem{Perspectives} J. P. Eisenstein, in \emph{Perspectives in Quantum
Hall Effects}, edited by S. Das Sarma and A. Piczuk (Wiley, New York, 1999).
J. Eisenstein \emph{et al.}, \prl{\textbf{62}}, 1540 (1989); L. W. Engel
\emph{et al.}, \prb{\textbf{45}}, 3418 (1992).

\bibitem{Smet} J. H. Smet \textit{et al.}, \prl{\textbf{86}}, 2412 (2001);
J.H. Smet, \emph{et al.}, Nature {\textbf{415}}, 281 (2002).




\bibitem{Kukushkin} I.V. Kukushkin \textit{et al.,} \prl{\textbf{82}}, 3665
(1999); I.V. Kukushkin \textit{et al.,} \prl{\textbf{85}}, 3688 (2000).

\bibitem{Chughtai2002} R. Chughtai \emph{et al.}, \prb{\textbf{65}},
161305(R) (2002).

\bibitem{Aifer96} E.H. Aifer \emph{et al.}, \prl{\textbf{76}}, 680 (1996);
M.J. Manfra \textit{et al.}, \prb{54} R17327 (1996). 

\bibitem{Yusa2001} G. Yusa, H. Shtrikman, and I. Bar-Joseph, %
\prl{\textbf{87}}, 216402 (2001); K. Kheng \textit{et al.}, \prl{71} 1752
(1993); D. R. Yakovlev \textit{et al.,} \prl{79}, 3974, (1997).

\bibitem{Wojs} A. Wojs and P. Hawrylak, \prb{51}, 10 880 (1995); A. Wojs, J.
J. Quinn and P. Hawrylak  \prb{\textbf{62}}, 4630 (2000).


\bibitem{Groshaus2004} J.G. Groshaus \emph{et al.} \prl{\textbf{93}}, 96802
(2004).


\bibitem{Murthy2000} G. Murthy, \prl{\textbf{84}}, 350 (2000).

\bibitem{OStern2004} O. Stern, \emph{et al.}, \prb{\textbf{70}}, 75318
(2004).

\bibitem{Schulze2004b} F. Schulze-Wischeler \emph{et al.}, \prl{\textbf{93}}%
, 26801 (2004); F. Schulze-Wischeler \emph{et al.}, Int. J. Mod. Phys. B
\textbf{18}, 3857 (2004).

\bibitem{Schulze2004a} F. Schulze-Wischeler, E. Mariani, F. Hohls, and R. J.
Haug \prl{\textbf{92}}, 156401 (2004).

\bibitem{comment-Triplet} The X peak could be associated with the triplet
trion, where the second electron is weakly bound.

\bibitem{Comment nu>1} For $\nu \geq 1$, not all electrons are unpaired. We
assume that the number of doubly occupied orbitals is $N_{e}-N_{\phi }$, and
thus, $f_{\downarrow ,\uparrow }=[N_{\downarrow ,\uparrow }-(N_{e}-N_{\phi
})]/N_{\downarrow ,\uparrow }$. Consequently, the expression for $\mathcal{P}
$ for $\nu \geq 1$ from Eq. \ref{truePol} needs to be multiplied by $(2-\nu
)/\nu $.

\bibitem{Laughlin83} R. B. Laughlin, \prl{\textbf{50}}, 1395 (1983); Rev.
Mod. Phys. \textbf{71}, 863 (1999).

\bibitem{CFSkyrmions} R. K. Kamilla \emph{et al.}, Sol. Stat. Com. \textbf{99%
}, 289 (1996); D. R. Leadley \emph{et al.}, \prl{\textbf{79}}, 4246 (1997);
A. F. Dethlefsen \emph{et al.}, arXiv:cond-mat/0508393 (2005); A. F.
Dethlefsen \emph{et al.}, arXiv:cond-mat/0603455 (2005).


\bibitem{comment-variance} Ref \cite{Kukushkin} shows that the critical
field is sensitive to the effective width of a 2DES. See also: S. S. Mandal
and J. K. Jain, \prb{\textbf{64}}, 125310 (2001).

\bibitem{Eros} E. Mariani \emph{et al.}, \prb{\textbf{66}}, 241303(R)
(2002); K. Vyborny and D. Pfannkuche, Int. J. Mod. Phys B \textbf{18} 3871
(2004).



\bibitem{Jain89} J.K. Jain, \prl{\textbf{63}}, 199 (1989); B.I. Halperin
\emph{et al.}, \prb{\textbf{47}}, 7312 (1993); K. Park and J.K. Jain, %
\prl{\textbf{80}}, 4237 (1998).

\end{thebibliography}
\end{document}